\documentclass[conference]{IEEEtran}
\IEEEoverridecommandlockouts
\usepackage{cite}
\usepackage{amsmath,amssymb,amsfonts}
\usepackage{algorithmic}
\usepackage{graphicx}
\usepackage{textcomp}
\usepackage{xcolor}
\usepackage{array}
\usepackage{wrapfig}
\usepackage{makecell}
\newcolumntype{P}[1]{>{\centering\arraybackslash}p{#1}}
\definecolor{lightgray}{HTML}{F7F7F7}
\definecolor{customblack}{HTML}{A0A0A0}
\usepackage{graphicx}
\usepackage{tcolorbox}
\usepackage{textcomp}
\usepackage{xcolor}
\usepackage{mathrsfs}
\usepackage{amsmath}
\usepackage{amsfonts}
\usepackage{amssymb}
\usepackage{url}
\usepackage{booktabs}
\usepackage[margin=1in]{geometry}
\usepackage{multirow}
\usepackage{float}
\usepackage{threeparttable}
\usepackage{afterpage}
\usepackage{subcaption}
\def\BibTeX{{\rm B\kern-.05em{\sc i\kern-.025em b}\kern-.08em
    T\kern-.1667em\lower.7ex\hbox{E}\kern-.125emX}}
\begin{document}


\title{``Energon": Unveiling Transformers from GPU Power and Thermal Side-Channels\\}



\author{\IEEEauthorblockN{Arunava Chaudhuri\IEEEauthorrefmark{1}, Shubhi Shukla\IEEEauthorrefmark{2}, Sarani Bhattacharya\IEEEauthorrefmark{1}, and Debdeep Mukhopadhyay\IEEEauthorrefmark{1}}
\IEEEauthorblockA{\IEEEauthorrefmark{1}\textit{Department of Computer Science and Engineering}, 
\textit{Indian Institute of Technology, Kharagpur}, Kharagpur, India \\
\IEEEauthorblockA{\IEEEauthorrefmark{2}\textit{Centre for Computational and Data Sciences}, 
\textit{Indian Institute of Technology, Kharagpur}, Kharagpur, India \\
\{arunavachaudhuri392, shuklashubhi6, bhattacharya.sarani.iitkgp, debdeep.mukhopadhyay\}@gmail.com}}
\textbf{Accepted at ICCAD 2025} 
}


\maketitle

\begin{abstract}

Transformers have become the backbone of many Machine Learning (ML) applications, including language translation, summarization, and computer vision. As these models are increasingly deployed in shared Graphics Processing Unit (GPU) environments via Machine Learning as a Service (MLaaS), concerns around their security grow. In particular, the risk of side-channel attacks that reveal architectural details without physical access remains underexplored, despite the high value of the proprietary models they target. This work to the best of our knowledge is the first to investigate GPU power and thermal fluctuations as side-channels and further exploit them to extract information from pre-trained transformer models. 
The proposed analysis shows how these side channels can be exploited at user-privilege to reveal critical architectural details such as encoder/decoder layer and attention head for both language and vision transformers. We demonstrate the practical impact by evaluating multiple language and vision pre-trained transformers which are publicly available. Through extensive experimental evaluations, we demonstrate that the attack model achieves a high accuracy of over 89\% on average for model family identification and 100\% for hyperparameter classification, in both single-process as well as noisy multi-process scenarios. Moreover, by leveraging the extracted architectural information, we demonstrate highly effective black-box transfer adversarial attacks with an average success rate exceeding 93\%, underscoring the security risks posed by GPU side-channel leakage in deployed transformer models.


\end{abstract}

\begin{IEEEkeywords}
side-channel, transformer, model stealing, GPU.
\end{IEEEkeywords}

\section{Introduction}

Transformer-based models like  Bidirectional Encoder Representations from Transformers (BERT), Large Language Models (LLMs), and Vision Transformers (ViTs) have significantly advanced natural language processing (NLP) and computer vision by capturing complex patterns in large datasets.Their high computational demands have driven companies like NVIDIA to develop specialized GPUs, with the growing reliance on such hardware reflected in NVIDIA’s rising stock prices. Despite this growing dependence on transformers and GPUs, the security of both against side-channel attacks remains limited and largely unexplored. In particular, there is currently no work examining the potential for model extraction or model-stealing attacks on transformers executed on GPUs through side-channel vulnerabilities, highlighting a critical gap in security research as these models continue to proliferate. 

\textit{Model extraction} or \textit{model-stealing} attacks aim to replicate or gain insight into a target deep learning model’s architecture or parameters without having direct access to it. Broadly, two types of model-stealing attacks have been studied. The first is the \textit{query-based model-stealing attack}, where attackers exploit a model’s prediction Application Programming Interface (API) to clone or replicate it, without accessing its parameters or training data. Query-based attacks have exposed vulnerabilities in models like DNNs and CNNs \cite{DBLP:journals/corr/abs-1912-07721,9,16,17}, and recent work has even extracted embedding layers from black-box transformer models such as OpenAI’s ChatGPT using API access alone~\cite{DBLP:conf/icml/CarliniPD0HCLJN24}.

The second type of attack, \textit{side-channel-based model-stealing}, exploits leakages such as power, thermal, electromagnetic emissions, and microarchitectural behaviors (e.g., cache accesses, branch misses) to infer model architecture. Unlike query-based methods, these attacks leverage shared hardware resources without directly interacting with the model API. Side-channel attacks use techniques which include cache-based channels \cite{56,10,24}, and physical leakages like power \cite{57}, thermal \cite{29}, electromagnetic emanations \cite{2,4,58}, and off-chip memory access \cite{13}. Prior GPU-based efforts have focused on extracting DNN/CNN architectures using CUPTI counters~\cite{DBLP:conf/dsn/WeiZZLF20} (currently not accessible) and resource-tracking APIs via CUDA-based spy applications~\cite{28}. \textit{However, no prior work has explored GPU side-channels specifically for transformer model computations.}

\textit{In this work, we address this gap by exploring GPU side-channels for transformer models, focusing specifically on power and thermal channels that remain accessible without special privileges and cannot be virtualized, even within virtualized GPU environments.} Our work specifically targets \textit{NVIDIA GPUs}, which currently holds 88\% of the GPU market share, though the approach can readily be extended to other GPUs. In the literature several side-channel attacks are performed on older generation of NVIDIA GPUs, including power and timing side-channel  attack on Kepler architecture\cite{7357115, 10.1145/3060403.3060462}, as well as timing side-channel attack on Volta\cite{9367179, 8852671} and Maxwell architecture\cite{10.1145/3243734.3243831}. However, post-Pascal generations of NVIDIA GPUs have undergone significant architectural changes, most importantly with the introduction of Multi-Instance GPU (MIG) (discussed in Section~\ref{sec:architecture_extraction}) technology in the Ampere generation for production-grade GPUs, enables better resource distribution and improved utilization in cloud environments.
Furthermore, there are not much documented side-channel experiments on post-Pascal generations of NVIDIA GPUs in the literature. \textit{In our experiment, we focus on newer generation of NVIDIA GPUs, specifically those based on the Turing and Ampere architectures.}

Despite advancements in underlying hardware of newer generation of NVIDIA GPUs (Turing, Ampere, Ada Lovelace), to the best of our knowledge, no GPU manufacturer offers protections against these side-channels. Our initial experiments show that transformers produce a distinctive \textit{staircase-like} pattern in side-channel traces, far more distinctive than those in earlier DNNs like CNNs. This pattern enables accurate inference of encoder-decoder layers from a single trace. We train CNNs on these side-channels to predict transformer's architectural details, achieving 95.00\% and 91.25\% accuracy in classifying language and vision model families, respectively. 
%
In summary, the contributions of this work are as follows:

\begin{itemize}
    
    \item We identify power and thermal traces as potential side-channels that reveal architectural details of transformer models and develop CNN-based predictors to infer attributes like encoder-decoder layers and attention heads.
    
    \item Moreover, to showcase the broad applicability of the observed GPU side-channels, we develop predictive models to identify the architectural families of 8 publicly available pre-trained language models and 8 vision models.

    \item We evaluate our attack against a noisy, multi-process environment and still achieve over 89\% accuracy in predicting black-box transformer model's parameters.
    
    \item We demonstrate that partial architectural leakage enables building a substitute language transformer model with a BERT Score over 93, and achieves 93\% success in black-box transfer attacks on Vision Transformers.
\end{itemize}

\textbf{Responsible Disclosure:} We have responsibly disclosed the power and
thermal side-channel vulnerability to NVIDIA. NVIDIA offered to acknowledge us on product security page stating that telemetry should be disabled for more security-sensitive use cases such as Confidential Compute (CC) and MIG, and mentioned they are cautiously evaluating how much more telemetry can be safely exposed. They also approved public disclosure of our findings.

\par
This paper is organized as follows: Section~\ref{sec:background} presents background on transformer models. Section~\ref{sec:side-channel_Leakages} analyzes GPU power and thermal side-channels during transformer inference. Section~\ref{sec:architecture_extraction} outlines our methodology for extracting architectural information and presents experimental results. Sections~\ref{sec:attack} and~\ref{sec:discussion} cover a black-box transfer adversarial attack and discuss about some of the related artifacts respectively. Finally, Section~\ref{sec:mitigation_conclusion} concludes with possible mitigations.

\section{Background on Transformers}
\label{sec:background}
Introduced by Google Brain in their paper \textit{Attention Is All You Need} \cite{1}, the transformer has become a state-of-the-art approach in natural language processing, surpassing earlier models like recurrent neural networks (RNNs) and long short-term memory networks (LSTMs). Unlike these models, which struggle with long sequences and slow training speeds, the transformer replaces recurrence with a fully attention-based mechanism, enabling efficient parallel processing of entire sequences. In the following, we briefly discuss about its core architectural modules.
\par


\textbf{Transformer Encoder:} 
In a transformer, the encoder processes the input sequence to generate a representation that captures contextual relationships among tokens. It consists of multiple layers stacked together, each receiving embedded input plus positional embeddings to retain token positions. Each layer includes a multi-head attention mechanism and a feed-forward network, both with residual connections and followed by layer normalization. In the multi-head attention layer, attention scores are calculated using query, key, and value vectors from input embeddings. The query-key dot product, passed through a softmax, yields attention outputs multiplied by the value vector and sent to the feed-forward network. The final encoder output is passed to the decoder layers.\par

\textbf{Transformer Decoder:} 
The decoder generates the output sequence using the encoder’s representation and previously generated tokens. Each decoder layer mirrors the encoder’s structure but includes masked self-attention to prevent future tokens from influencing the current token. The final layer outputs a translated word, which is added to the decoder input to continue the process. This cycle repeats until an end-of-sequence token is predicted, enabling it to handle sequence-to-sequence tasks.\par

\textbf{Self-Attention:} In addition to the encoder and decoder, the transformer model uses self-attention, allowing each token in a sequence to focus on others, capturing contextual relationships regardless of position. Multi-head attention achieves this, with each head learning different aspects of token relationships. In the encoder, self-attention gives each token access to all tokens in the input, while in the decoder, it is masked to prevent future tokens from influencing the current token.

Each of the transformer's components impose distinct computational demands, shaping unique power and thermal profiles. We aim to exploit these patterns to uncover architectural insights via side-channel analysis.

\section{Power and Thermal Foot prints of Transformer}
\label{sec:side-channel_Leakages}

Power and thermal side-channels have been extensively used to extract secure information such as cryptographic keys \cite{c1}, plaintext data \cite{c2}, confidential algorithms \cite{c3}, and more from various computational systems. Recently, these side-channels have also been applied to deep learning models, revealing private information including architectural details \cite{c5}, model weights \cite{c6}, and training hyperparameters \cite{c7} when executed on CPUs and edge devices. Traditionally, side-channel attacks and model fingerprinting techniques have focused primarily on simpler DNN architectures like CNNs. However, as complex and widely-used transformer-based architectures emerge, they remain relatively unexplored in terms of security and side-channel vulnerabilities. This work aims to address this gap by investigating side-channel-based attacks specifically targeting transformer models, highlighting potential security risks. An important aspect of our work is the application of side-channel analysis on GPUs, as transformers are frequently hosted on large data center GPUs to enable faster computation and support larger model sizes.

\begin{figure}[ht]
    \vspace{-3mm}
    \centering
    \begin{subfigure}[b]{0.24\textwidth} 
        \centering
        \includegraphics[width=\textwidth]{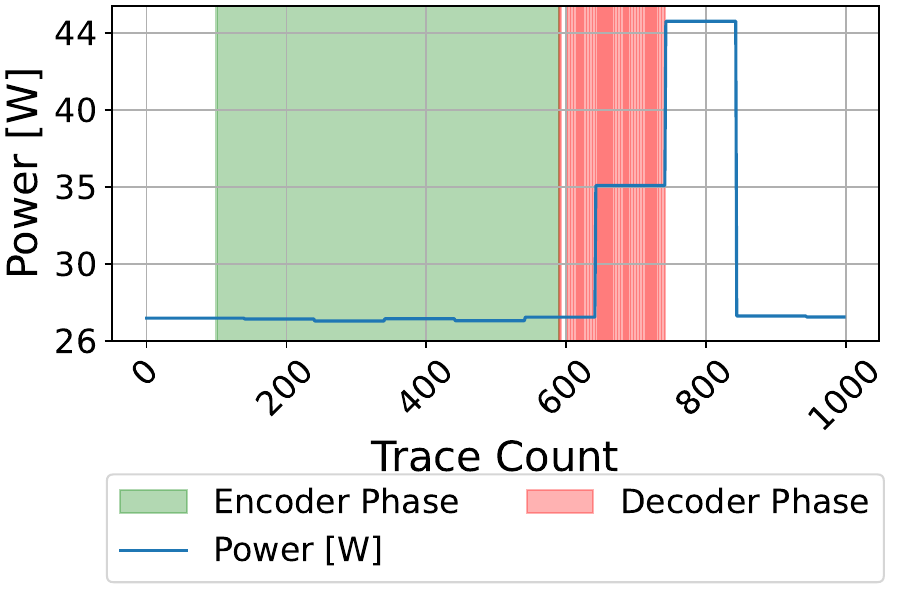} 
        \caption{Transformer Model}
        \label{fig:Figure1}
    \end{subfigure}
    \begin{subfigure}[b]{0.24\textwidth} 
        \centering
        \includegraphics[width=\textwidth]{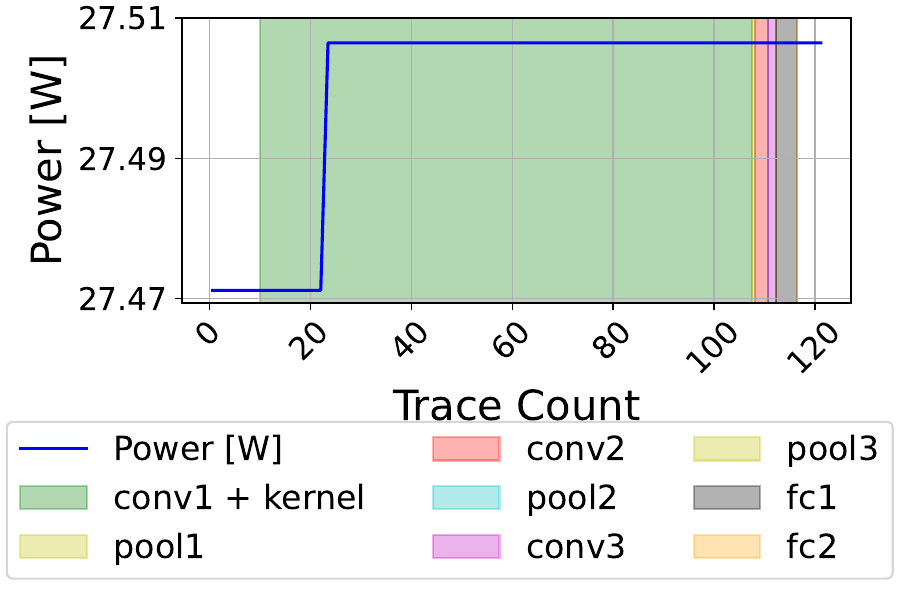} 
        \caption{CNN Model}
        \label{fig:custom_cnn}
    \end{subfigure}
    \caption{Power Trace of a custom CNN and transformer model with one encoder/decoder layer and eight attention heads during single inference. \vspace{-4mm}}
    \label{fig:single_trace_power_comp}
\end{figure}

\begin{figure*}[ht]
    \centering
    \begin{subfigure}[b]{0.27\textwidth} 
        \centering
        \includegraphics[width=\textwidth]{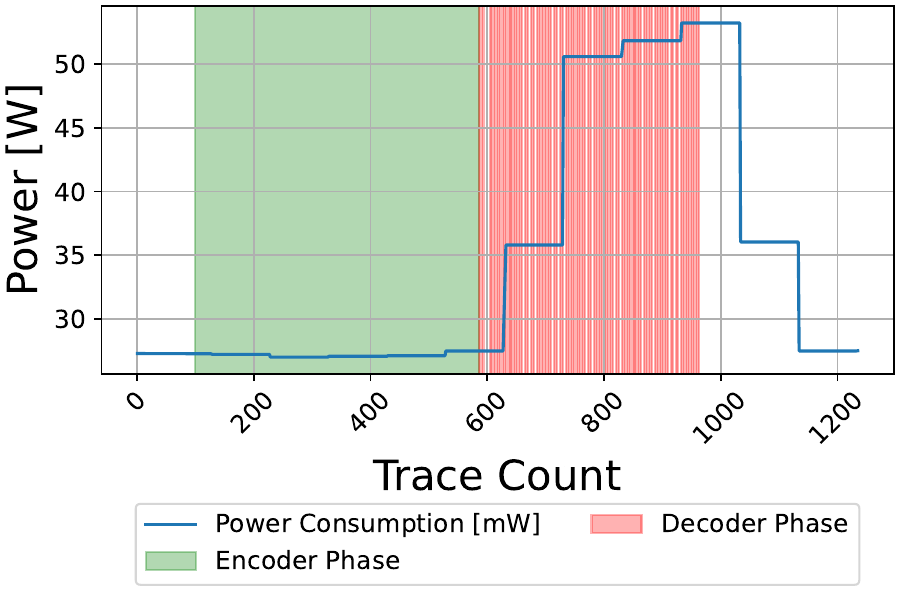} 
        \caption{Encoder/ Decoder: 3}
        \label{fig:encoder_decoder3}
    \end{subfigure}
    \hfill
    \begin{subfigure}[b]{0.27\textwidth} 
        \centering
        \includegraphics[width=\textwidth]{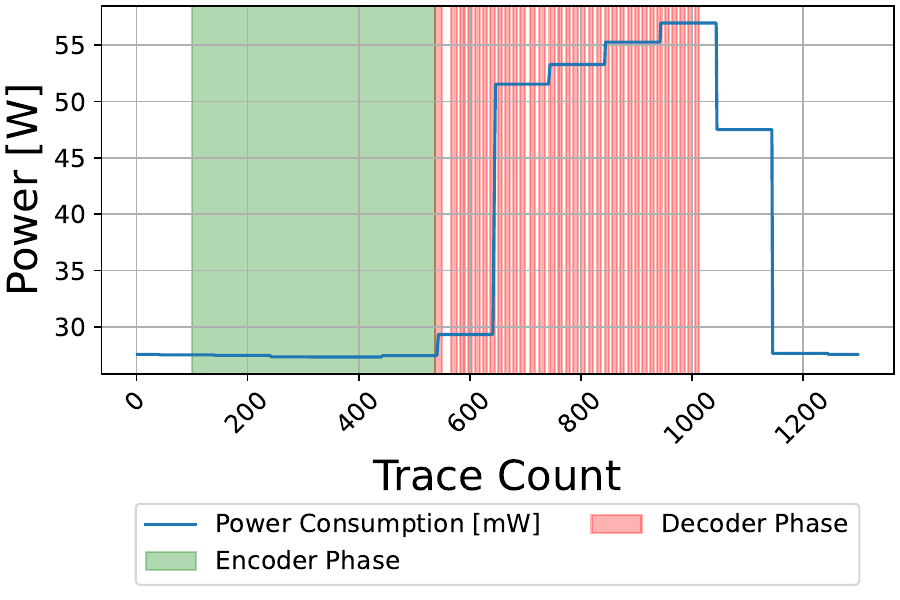} 
        \caption{ Encoder/ Decoder: 5}
        \label{fig:encoder_decoder5}
    \end{subfigure}
    \hfill
    \begin{subfigure}[b]{0.27\textwidth} 
        \centering
        \includegraphics[width=\textwidth]{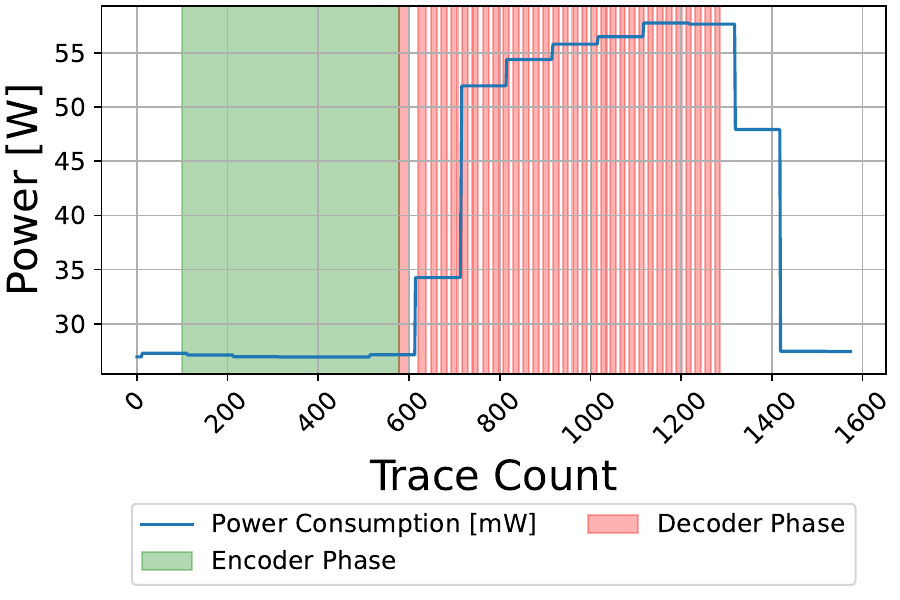} 
        \caption{Encoder/ Decoder: 8}
        \label{fig:encoder_decoder8}
    \end{subfigure}
    \caption{Power Trace of single transformer Model inference with varying encoder/decoder and eight attention heads. \vspace{-9mm}}
    \label{fig:multiple_single_trace_power}
\end{figure*}

\vspace{-1.5mm}
\subsection{\textbf{Power Gradient Analysis for Transformer Models}}
We aim to explore the impact of transformer model execution on GPU power consumption and as an initial step, we built a custom language transformer model. Our first model was trained with a configuration of one encoder and one decoder with eight attention heads. We set the embedding vector size to 512, and the maximum sentence length was limited to 200.  The power trace was collected with user privileges using \texttt{nvidia-smi} query with GPU temperature stabilized at 28$^{\circ}$C.  Our goal was to collect data at a sampling rate of 10 Hz or 100 Hz, but the recorded trace length was extremely short due to faster model inference, and no distinguishable trends were observed for transformer execution. To address this, we aimed to set the sampling rate at 1 Hz to increase the trace length but had to settle for 7 Hz, as it was the minimum achievable rate. Additionally, at high temperatures, the GPU consumes more power, leading to erroneous information for transformer model inference. To mitigate this, we collected all traces while keeping the GPU temperature stable and allowed the GPU to cool down before the next execution cycle begins, ensuring that previously collected traces do not influence subsequent ones. Consequently, the overall trace collection time was increased due to this additional waiting time required for the GPU to cool down.
This approach is effective as long as the GPU caches and memory are cleared of the prior process’s data elements. Such a methodology is commonly employed for reliable data collection in side channel analysis. In real-world scenarios  for similar analyses, additional GPU cooling would not be necessary, as traces are typically gathered from the same process running continuously for inference workloads.

Using the trained model, we performed an inference on a consumer-grade GPU (GTX 1660 Ti Mobile GPU) and plotted its power trace data, as shown in Figure \ref{fig:Figure1}. The first green-shaded zone in the figure represents power usage during encoder execution, while the consecutive red-colored slender areas indicate the decoder execution as it generates the translated output for each word in the input phrase. Here, we observe a staircase-like pattern emerging during transformer model inference. Notably, the power consumption pattern differs significantly between the encoder and decoder phases: during encoder execution, power consumption rises to 6W, whereas, during decoder execution, power demand increases to 20W. \textit{This happens because, during the execution of the encoder, it processes a fixed-length input sequence at a time to create the final attention matrix, which is then used as input to the decoder. This explains the initial small increase in the power graph before it stabilizes. In contrast, the decoder operates on a gradually increasing input sequence to generate each translated word. As the input size grows, the computation becomes more complex, leading to a steady step-wise increase in power consumption until the final translated word is generated.} For comparison, we also present power readings for a custom CNN model in Figure~\ref{fig:custom_cnn} and observe no distinctive pattern across its layers, unlike the unique patterns seen in transformers in Figure~\ref{fig:Figure1}. Instead, the CNN exhibits a steady power reading of approximately 27.5 W throughout execution. Consequently, in this work, we specifically target transformer architectures due to their uniquely observable patterns.

\begin{figure}
\centering
\includegraphics[width=6.5cm]{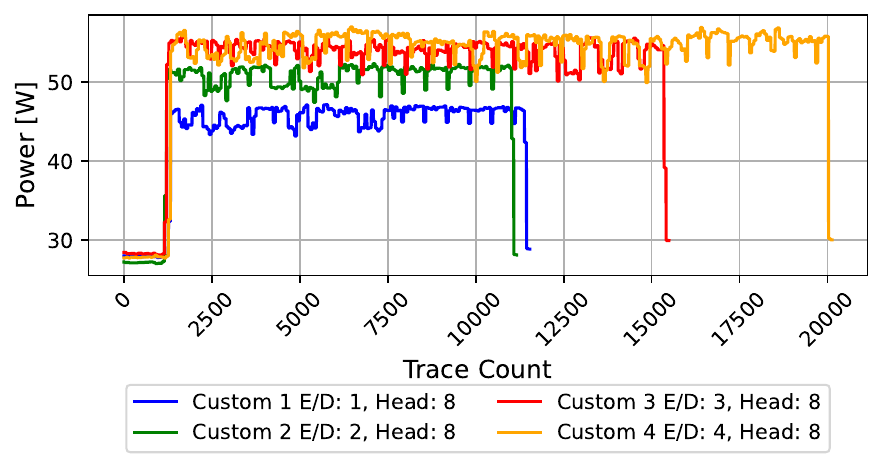}
   \caption{GPU Power Trace of Transformer Model during 100 Inferences on NVIDIA GTX 1660 Ti (E/D: Encoder/Decoder).\vspace{-6mm}}
\label{fig:Figure2}
\end{figure}

Following the observed results, we extended our analysis to larger transformer models to see if similar behavior could be replicated. We created and trained four additional transformer models with varying encoder-decoder configurations. Next, we collected power trace data for each model running on the same GPU, as shown in Figure \ref{fig:multiple_single_trace_power}. The power trace data reveals a clear relationship between power consumption and transformer size (number of encoder/decoder blocks). As the number of encoder/decoder blocks increases, so does the power consumption during execution. When the transformer's decoder becomes active during inference, power consumption increases to as much as 30 W across nearly all model inferences. Additionally, we observed that the number of staircase steps in the power trace grows proportionally with the number of encoder/decoder blocks in the transformer model.

\begin{figure*}[t]
    \setlength{\tabcolsep}{6pt} 
    \renewcommand{\arraystretch}{1.1} 
    \begin{minipage}[c]{0.62\textwidth}
        \centering
        \scriptsize 
        \centering 
        \captionof{table}{Popular Pre-Trained Language Transformer Configurations}
        \vspace{-0.2cm}
        \label{tab:transformer_config}
        \begin{tabular}{|c|c|c|c|c|}
            \hline
            \multirow{2}{*}{\textbf{Family}} & \multirow{2}{*}{\textbf{Transformer Name}} & \textbf{\#E/D} & \textbf{\#Attention} & \textbf{Embedding}\\ 
            &&& \textbf{Heads} & \textbf{Dimension}\\
            \hline
            T5 & t5-small & 6 / 6 & 8 & 512\\ \hline
            T5 & t5-base & 12 / 12 & 12 & 768\\ \hline
            T5 & t5-large & 24 / 24 & 16 & 1024\\ \hline
            T5 & t5-3b & 24 / 24 & 32 & 1024\\ \hline
            MarianMT & Helsinki-NLP/opus-mt-en-fr & 6 / 6 & 8 & 512\\ \hline
            META & facebook/nllb-200-distilled-600M & 12 / 12 & 16 & 1024\\ \hline
            META & facebook/nllb-200-distilled-1.3B & 24 / 24 & 16 & 1024\\ \hline
            Google & madlad400-3b-mt & 32 / 32 & 16 & 1024\\ \hline
        \end{tabular}
    \end{minipage}%
    \hfill
    \begin{minipage}[c]{0.36\textwidth}
        \centering
        \includegraphics[width=0.9\linewidth]{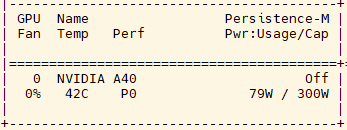}
        \captionof{figure}{ \texttt{nvidia-smi} query output displaying power and thermal data for NVIDIA A40 GPU.}
        \label{fig:Figure8}
    \end{minipage}
    \vspace{-4mm}
\end{figure*}

\vspace{-1.5mm}
\subsection{\textbf{Power Side-channel on Data Center GPUs}}
With above knowledge and understanding of transformer power trace characteristics, we extend our experiments on a larger scale with practical use cases using server-grade GPUs. For this, we replicate the scenario of transformer models running on a data center GPU, where multiple batch inferences are typically executed concurrently. To simulate this, we collect power trace data from our custom transformer models by performing 100 continuous inferences on each model. In Figure~\ref{fig:Figure2}, we observe a clear difference in power consumption across each configuration of the transformer model.\par

\begin{figure*}[ht]
    \centering
    \begin{subfigure}[b]{0.49\textwidth} 
        \centering
        \includegraphics[width=\textwidth]{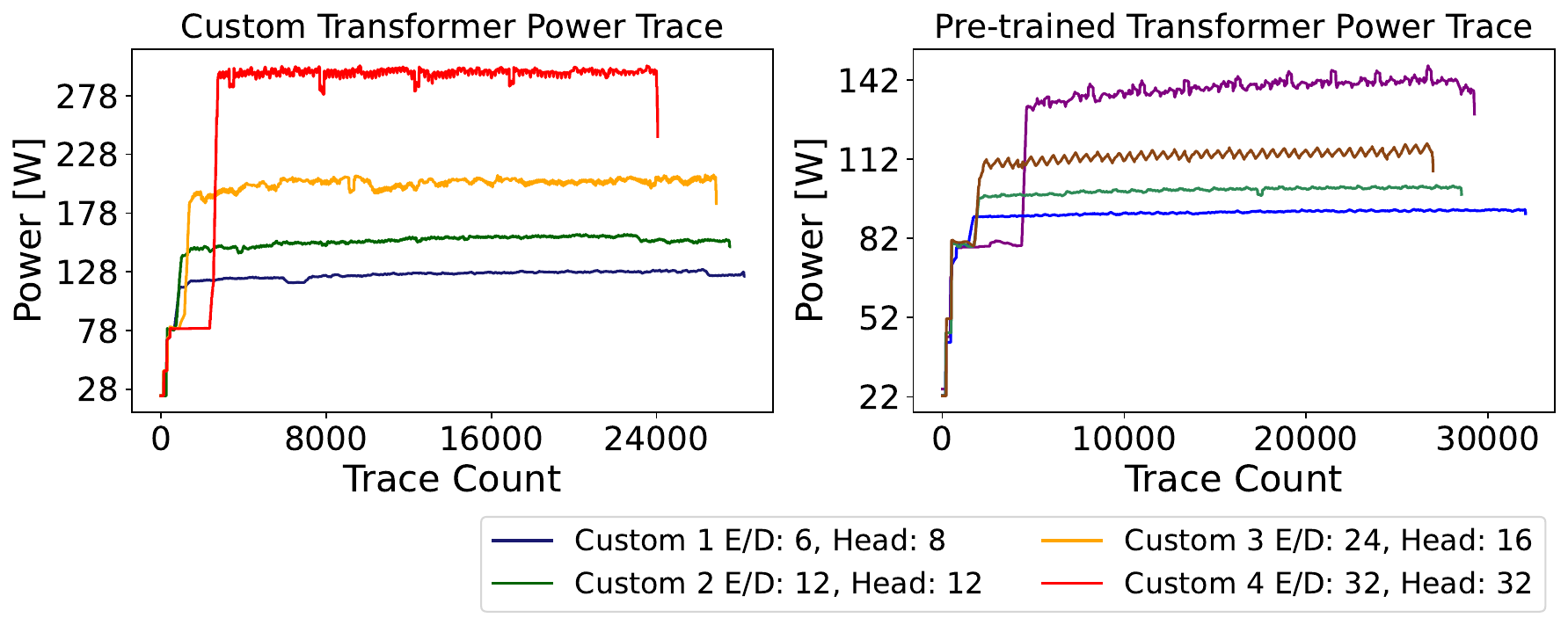} 
        \caption{Power Traces}
        \label{fig:Figure3}
    \end{subfigure}
    \begin{subfigure}[b]{0.49\textwidth} 
        \centering
        \includegraphics[width=\textwidth]{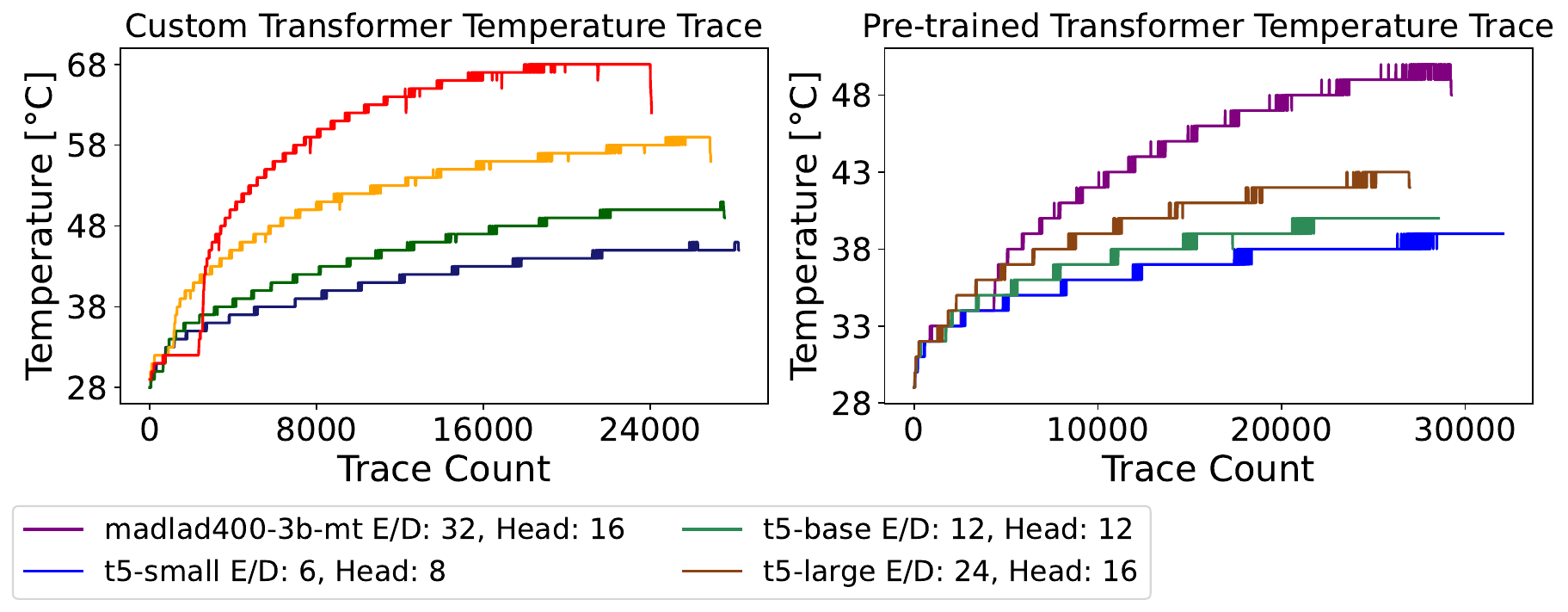} 
        \caption{Thermal Traces}
        \label{fig:Figure5}
    \end{subfigure}
    \caption{ Power and thermal traces of transformers on NVIDIA A40 during 120s inference (E/D: Encoder/Decoder)\vspace{-5mm}}
    \label{fig:multiple_single_trace_power2}
\end{figure*}

To further demonstrate the impact of this side-channel, we built and trained four custom models with configurations similar to those of popular pre-trained language models on Hugging Face, as shown in Table \ref{tab:transformer_config}. This time, we extended our observations to data center GPUs commonly used for running large transformer models in various AI tasks. Therefore, all subsequent experiments are conducted on the NVIDIA A40 GPU. We collected power trace data for 120 seconds while running inference with our previously trained custom models. In this setup, we observed a similar pattern in  Figure \ref{fig:Figure3} (Custom Transformer Power Trace) as in our earlier experiment (Figure \ref{fig:Figure2}), though power consumption differences are significantly higher, while differences between models remain apparent. Additionally, to gain insights into practical applications, we collected GPU power trace data while running inference on popular pre-trained models from Hugging Face. In Figure \ref{fig:Figure3} (Pre-trained Transformer Power Trace), we observe distinct power traces for each transformer model, depending on model size (number of encoder/decoder layers and attention heads). Each model exhibits a unique power consumption pattern that can aid in its identification when executing on GPUs. 

\vspace{-1.5mm}
\subsection{\textbf{Thermal Side-channel on Data Center GPUs}}
Given the observed variations in power consumption based on the size and complexity of transformer models on GPUs, we anticipate similar effects on other side-channel metrics, such as thermal data. To verify this, we conducted an experiment similar to the previous one, collecting temperature data over 120 seconds during transformer model inference. In the resulting temperature trace graph (Figure \ref{fig:Figure5}), we observe a stepwise, gradual increase in temperature throughout the inference duration. This indicates that the rate of temperature rise correlates with the transformer model’s size on the GPU, similar to our findings from the power traces.

\begin{tcolorbox}[
    colback=lightgray,   
    colframe=customblack,       
    arc=2mm,             
    boxrule=1pt,          
    left=1pt,   
    right=1pt,  
    top=1pt,    
    bottom=1pt, 
]
\textit{\textbf{Takeaway :} Based on the above observations, we conclude that each transformer model exhibits distinct temperature and power consumption patterns, which can uniquely differentiate them from other models running on consumer or production-grade GPUs.}
\end{tcolorbox}

\section{Extracting Transformer Architecture Using GPU Side-channels}
\label{sec:architecture_extraction}
Building on observations from the previous section, we now aim to leverage power and thermal side-channels to extract architectural details of well-known transformer models, including both language and vision models. To proceed, we first define our threat model.

\vspace{-1.5mm}
\subsection{\textbf{Threat Model:}} 

We consider a scenario with multiple users on a remote cloud server. To provide single-GPU access to multiple clients, cloud providers use GPU virtualization technologies like NVIDIA vCS combined with Multi-Instance GPU (MIG) backend technology, in contrast to traditional time-shared setups. Modern cloud providers like AWS and Google Cloud offer dedicated GPU instances per VM, allowing multiple users to run tasks independently. According to NVIDIA’s MIG documentation, each GPU instance has isolated compute, memory, and bandwidth resources, minimizing interference and context-switching issues. In addition, users can access power and thermal metrics via tools like \textit{nvidia-smi} and \textit{pynvml} without root privilege. These metrics are directly obtained from physical GPU sensors and cannot be virtualized, even with MIG.

In this setup, a victim runs an unknown model on one virtualized GPU instance, while an adversary in a separate VM tries to infer victim model's architecture through other GPU instances. Below, we outline the adversary’s capabilities and objectives in detail:

\noindent \textbf{Adversary’s capabilities :} The adversary can continuously monitor thermal and power data for its assigned virtual GPU using tools like \texttt{nvidia-smi} (Figure \ref{fig:Figure8}) or the Python library \texttt{pynvml}. {\em Notably, the adversary does not need sudo/root privileges to access this information.}

\noindent \textbf{Adversary’s Objective :} The adversary’s goal is to extract architectural information about the transformer model running on the victim’s VM by monitoring power and thermal data from their own VM using the tools mentioned above.

\begin{figure*}
    \centering
    \begin{subfigure}[b]{0.20\textwidth} 
        \centering
        \includegraphics[width=\textwidth]{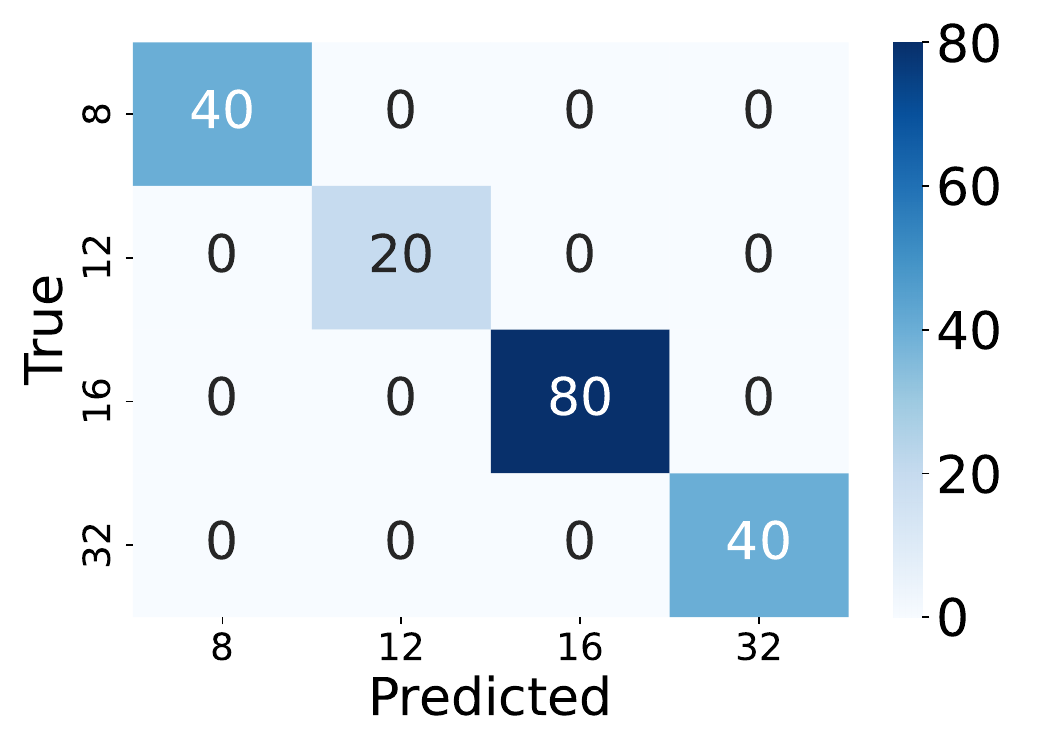} 
        \caption{Attention head prediction}
        \label{fig:custom_head}
    \end{subfigure}
    \hfill
    \begin{subfigure}[b]{0.20\textwidth} 
        \centering
        \includegraphics[width=\textwidth]{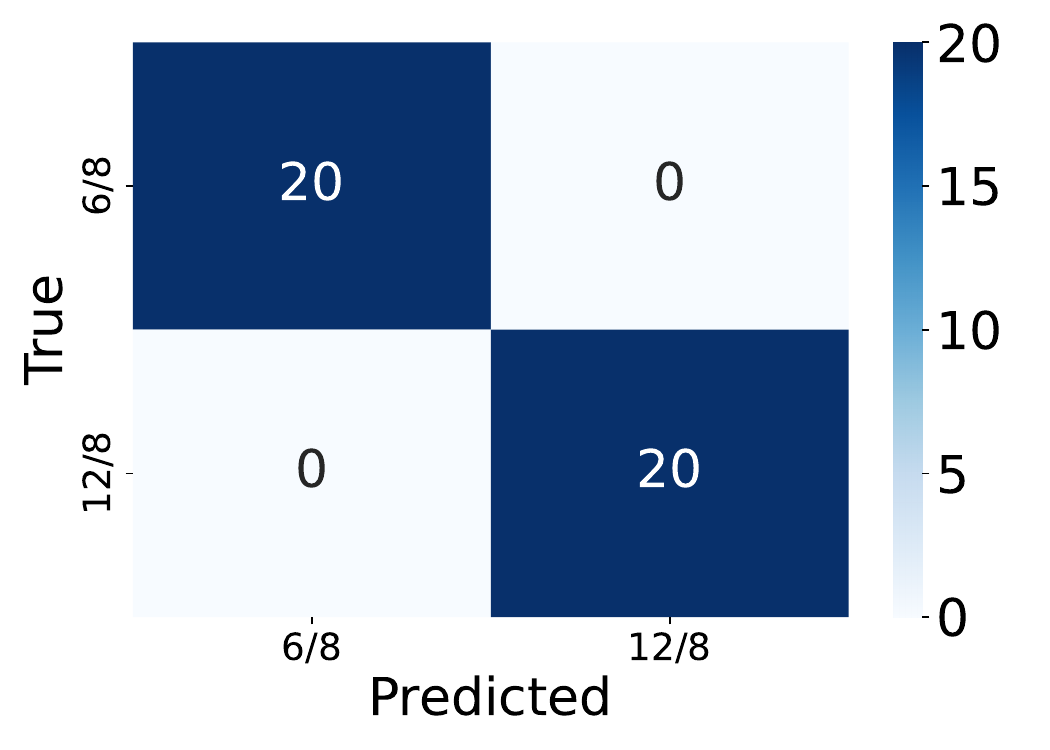} 
        \caption{Attention head 8}
        \label{fig:custom_8}
    \end{subfigure}
    \hfill
    \begin{subfigure}[b]{0.20\textwidth} 
        \centering
        \includegraphics[width=\textwidth]{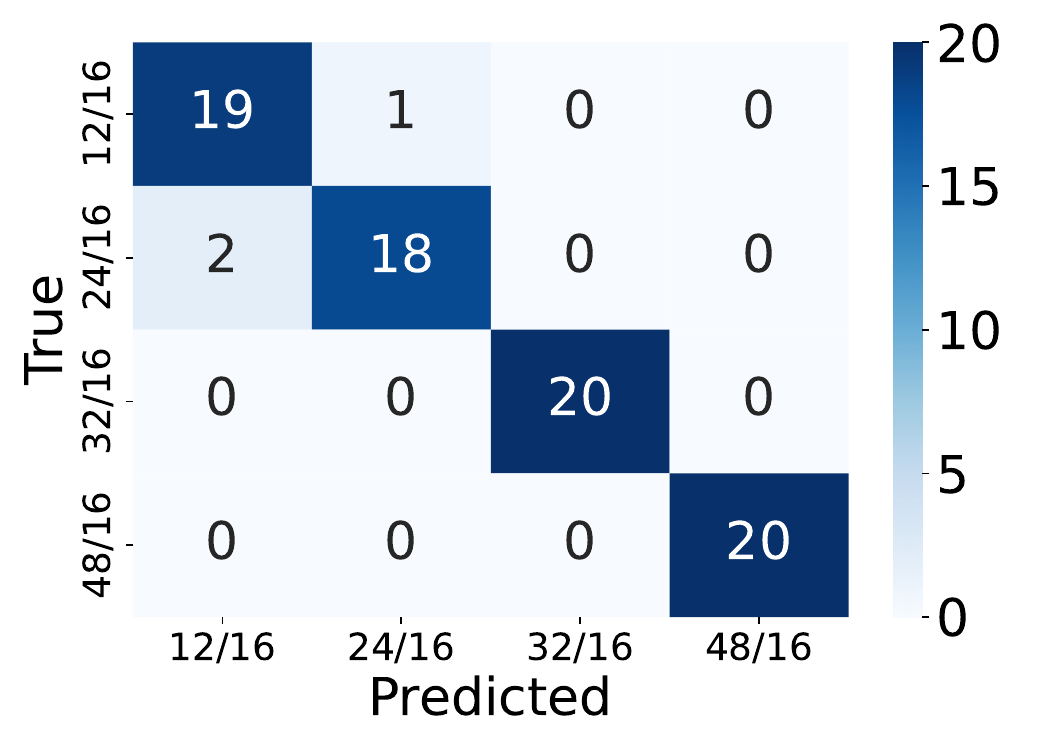} 
        \caption{Attention head 16}
        \label{fig:custom_16}
    \end{subfigure}
    \hfill
    \begin{subfigure}[b]{0.20\textwidth} 
        \centering
        \includegraphics[width=\textwidth]{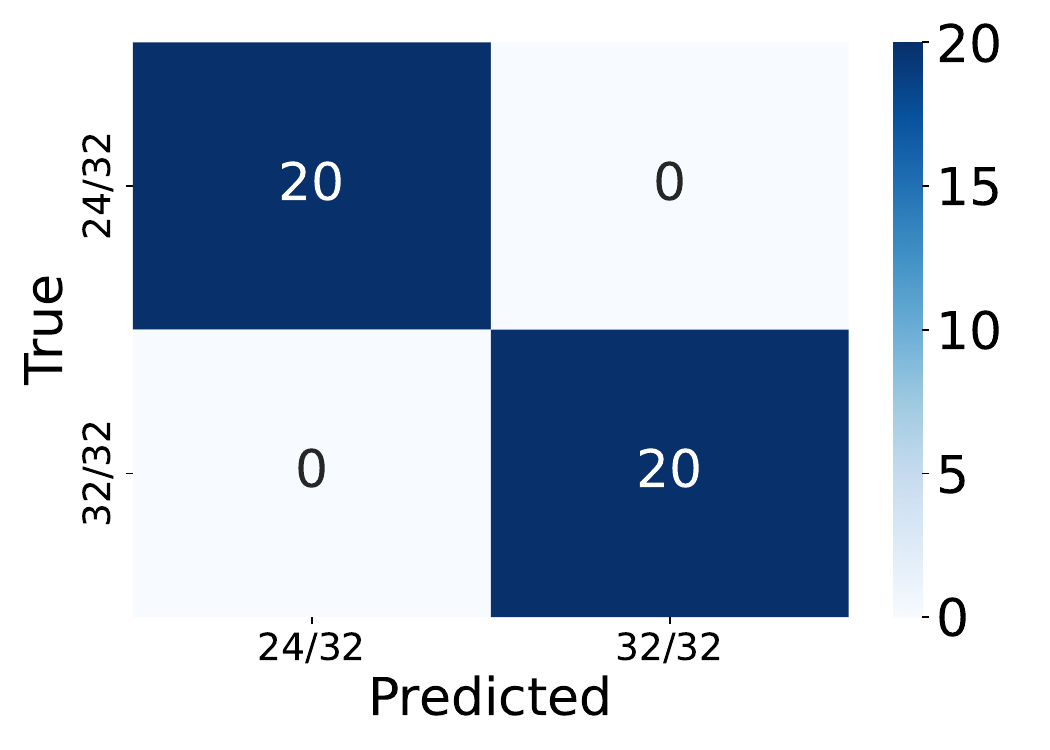} 
        \caption{Attention head 32}
        \label{fig:custom_32}
    \end{subfigure}
    \caption{ Confusion matrices for attention head and encoder/decoder layer prediction for attention head families.\vspace{-2.5mm}}
    \label{fig:custom_model_results}
\end{figure*}

\begin{table*}[ht]
    \centering
    \caption{Popular Pre-Trained Vision Transformer Basic Configuration (E/D: Encoder/Decoder)}
    \begin{tabular}{|c|c|c|c|c|c|}
    \hline
    \textbf{Family} & \textbf{Transformer Name} & \textbf{\#E/D} & \textbf{\#Attention Heads} & \textbf{Embedding Dimension} & \textbf{Input Image Size} \\ \hline
    Google & vit-base-patch16-224 & 12 & 12 & 768 & \textit{$224 \times 224$}\\ \hline
    Google & vit-large-patch16-225 & 24 & 16 & 1024 & \textit{$224 \times 224$}\\ \hline
    Apple & mobilevit-small & 12 & 4 & 384 & \textit{$256 \times 256$}\\ \hline
    META & deit-tiny-distilled-patch16-224 & 12 & 3 & 192 & \textit{$224 \times 224$}\\ \hline
    META & deit-small-distilled-patch16-224 & 12 & 6 & 384 & \textit{$224 \times 224$}\\ \hline
    META & deit-base-distilled-patch16-224 & 12 & 12 & 768 & \textit{$224 \times 224$}\\ \hline
    Microsoft &	swin-tiny-patch4-window7-224 &	12 & 3 & 96 & \textit{$224 \times 224$}\\ \hline
    Microsoft &	swin-base-patch4-window7-224 &	12 & 12 & 768 & \textit{$224 \times 224$}\\ \hline
    \end{tabular}
    \label{tab:vision_transformer_config}
    \vspace{-4mm}
\end{table*}

\vspace{-1.5mm}
\subsection{\textbf{Transformer Architecture Extraction}} 
We now aim to demonstrate the process of building a prediction model to extract key architectural parameters of transformer models based on our defined threat model. We present results using our custom models, pre-trained language and vision transformers from Hugging Face, aiming to extract key architectural details such as the number of attention heads, and encoder/decoder layers. Both power and thermal traces are utilized as inputs to the prediction model. These traces were collected over 120 seconds at a sampling rate of 7 Hz, with an initial GPU base temperature of 28$^{\circ}$C. In total, we gathered 100 traces for each model in our work, splitting the data into an 80:20 ratio for training and testing the prediction model, and utilized stratified \textit{K-fold} cross-validation for building and validating the model. For clarity, transformer configurations are denoted as {\sf X/Y}, where {\sf X} represents the number of encoder-decoder layers and {\sf Y} the number of attention heads.

\subsubsection{\textbf{Custom Language Transformer}}
We created and trained a total of nine custom transformer models with the following configurations: 6/8, 12/8, 12/12, 12/16, 24/16, 32/16, 48/16, 24/32 and 32/32. Our objective is to build machine learning models capable of predicting hyperparameter
of an unknown transformer model, using power and thermal traces as inputs. This is done in two steps: first classifying number of attention heads, followed by predicting encoder/decoder counts.
To achieve this, we built a prediction model consisting of three convolutional layers with sizes 32, 16, and 8, each followed by batch normalization and \textit{ReLU} activation. The model also includes two max-pooling layers and two fully connected layers with a \textit{softmax} activation function. The \textit{Adam} optimizer with a learning rate of $0.00001$ was used for training. The first model was trained for attention head classification, achieving 100\% accuracy (Figure \ref{fig:custom_head}) on test sets using both thermal and power traces. Subsequently, the prediction model further sub-categorized each attention head family based on the number of layers. The attention head count of 8 (two sub-classes) and 32 (two sub-classes) were classified with 100\% accuracy (Figure \ref{fig:custom_8} and \ref{fig:custom_32}), while the 16 attention head model (four sub-classes) achieved 96.25\% accuracy (Figure \ref{fig:custom_16}).


\subsubsection{\textbf{Pre-trained Language Model}}
Additionally, we present power/thermal-based prediction results for four pre-trained language transformer families: \textit{T5, Google, MarianMT}, and \textit{META}. As shown in Table \ref{tab:transformer_config}, \textit{T5} family includes four models, while \textit{Meta} has two, and \textit{Google} and \textit{MarianMT} each contributes one model. Unlike custom transformers, we consider eight pre-trained models across four research labs. Consequently, our first objective is to identify the target model's family, followed by classification of architectural details such as attention heads and encoder/decoder layers. This is feasible as models within the same family exhibit similar power and thermal patterns due to shared architectures and training methodologies.
Following this plan, we constructed our first prediction model to classify the families of pre-trained models using the same CNN architecture as before, achieving up to 95\% accuracy and an average of 89\% across five folds. To identify additional architectural details, we focus on two families: \textit{T5} and \textit{Meta}, developing separate prediction models for each of them.

\begin{table}[ht]
    \centering
    \vspace{-1.5mm}
    \caption{Pre-Trained Language Transformer Architecture Prediction Accuracy (Thermal \& Power Traces)}
    \resizebox{\columnwidth}{!}{%
    \begin{tabular}{|c|c|c|c|}
    \hline
    \textbf{Transformer Family} & \textbf{Prediction Criteria} & \textbf{Max Acc.(\%)} & \textbf{Avg. Acc.(\%)} \\ \hline
    All Models & Root Family & 95\% & 89\%\\ \hline
    META & Encoder/Decoder & 100\%  & 100\%\\ \hline
    T5 & Attention Head & 100\% & 100\%\\ \hline
    \end{tabular}
    }
    \label{tab:pre-trained_model_results}
    \vspace{-2mm}
\end{table}
As shown in Table~\ref{tab:transformer_config}, the \textit{T5} family models differ only in attention heads and layers, so we train a model to distinguish between the four variants based on attention heads. 
In contrast, the models from \textit{Meta} family share the same number of attention heads but vary in the number of encoder/decoder and total layers. Therefore, we build a predictive model for encoder/decoder layers to differentiate between the two possible models in this family. As shown in Table~\ref{tab:pre-trained_model_results}, both the \textit{T5} and \textit{Meta} prediction models achieve 100\% accuracy in identifying attention heads and encoder/decoder layers, respectively, from the test samples.\par

\subsubsection{\textbf{Pre-trained ViT Model}}
We further investigate popular ViT models from several renowned research labs on Hugging Face (refer to Table \ref{tab:vision_transformer_config}). Unlike language transformers, which feature an encoder-decoder structure, ViT models utilize a simpler encoder-only architecture. These models are also smaller and less complex than other transformer variants, allowing efficient execution on lower-end consumer GPUs.
Most ViT models consist of 12 layers, with 3 to 16 attention heads, 
and variable embedding dimensions depending on the model’s size. Nearly all the models studied accept input images of size \textit{$224 \times 224$}. For our experiments, the input to these transformers is derived from the CIFAR-10 dataset\cite{cifar-10}, which includes 60,000 color images of size \textit{$32 \times 32$}, distributed across 10 classes.\par

\begin{table}[ht]
\centering
\vspace{-1.5mm}
\caption{Pre-Trained Vision Transformer's Architecture Prediction Accuracy}
\label{tab:pre-trained_vision_model_results}
\resizebox{\columnwidth}{!}{%
\begin{tabular}{|c|c|c|c|}
\hline
\textbf{Transformer Family} & \textbf{Prediction Criteria} & \textbf{Max Acc.(\%)} & \textbf{Avg. Acc.(\%)} \\ \hline
All Models & Root Family & 91.25\% & 84.75\% \\ \hline
META & Attention Head & 100\% & 100\% \\ \hline
Google & Attention Head & 100\% & 100\% \\ \hline
Microsoft & Attention Head & 100\% & 100\% \\ \hline
\end{tabular}%
}
\vspace{-2mm}
\end{table}

Architectural differences in ViT models are shaped by attention heads, embedding dimensions, optimization methods, and tokenization strategies. We focus on four model families: \textit{Apple, Google, Facebook}, and \textit{Microsoft}. ViT models from \textit{Google} vary in the number of attention heads, encoders/decoders, and embedding dimensions, while those from \textit{Facebook} and \textit{Microsoft} differ mainly in attention heads and embedding sizes. 
Based on these distinctions, our initial prediction model classifies the models by their respective families followed by family-specific models that predict attention head counts. 
Following the same approach, 
our first prediction model (using the same CNN architecture as for custom transformer model prediction) for ViT models achieved a maximum accuracy of 91.25\% with an average of 84.75\% across five folds. 
Additionally, as shown in Table \ref{tab:pre-trained_vision_model_results}, prediction models for remaining three families achieve 100\% accuracy in predicting attention head counts using both thermal and power traces.

\subsection{\textbf{Model Extraction in Noisy Environment:}}
We have further evaluated the practicality of our approach for real-world scenarios where multiple processes run concurrently on the GPU. 
\subsubsection{\textbf{Model Extraction Against Variable Number of Background Process}}
\begin{figure}[ht]
\centering
\includegraphics[width=7cm]{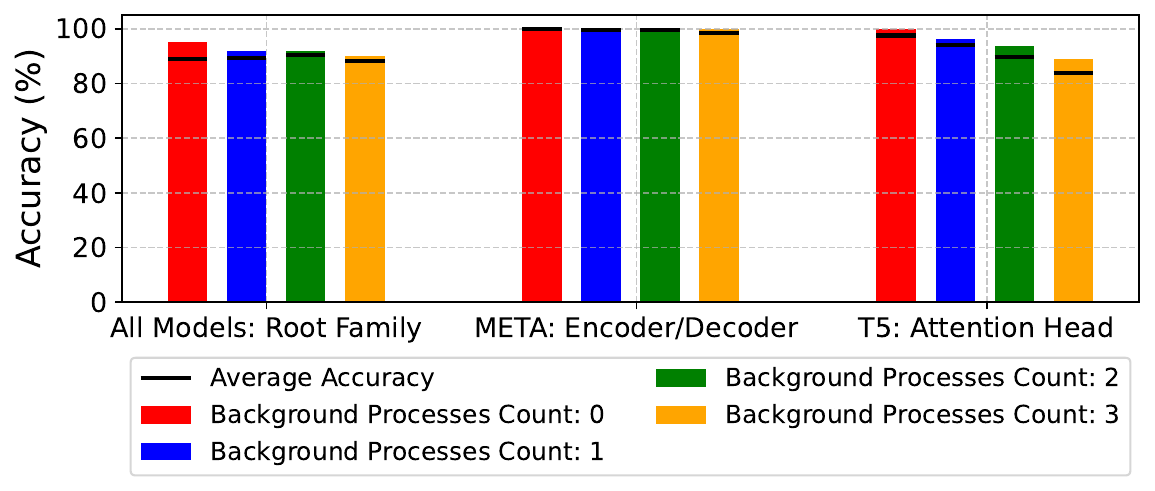}
   \caption{Prediction Model performance with increasing number of background process: Maximum Accuracy with Average as Dashed Line.\vspace{-2mm}}
\label{fig:Figure9}
\end{figure}

In server-grade GPUs, we focus on scenarios where a cloud GPU is partitioned into more than one independent GPU instances, with multiple processes are running independently within each instances. Under this setup, we executed transformer model inference and matrix multiplication in parallel and collected corresponding power and thermal traces. As shown in Figure~\ref{fig:Figure9}, despite the added workload noise, our classification model achieved an average accuracy of 89.25\% for identifying the transformer model family, and 92\% and 87\% accuracy in distinguishing between \textit{META} and \textit{T5} architectures—closely aligning with results from the ideal scenario. We further extended our experiments to include three and four concurrent processes, observing only a minimal drop in performance across tasks such as model family and attention head identification. Still, these results confirm that even in complex, noisy environments with multiple simultaneous processes, our attack model can reliably extract architectural details of black-box transformer models.

\vspace{-0.8mm}

\subsubsection{\textbf{Model Extraction Against Different Categories of Background Process}} 
\begin{figure}[ht]
\centering
\includegraphics[width=7cm]{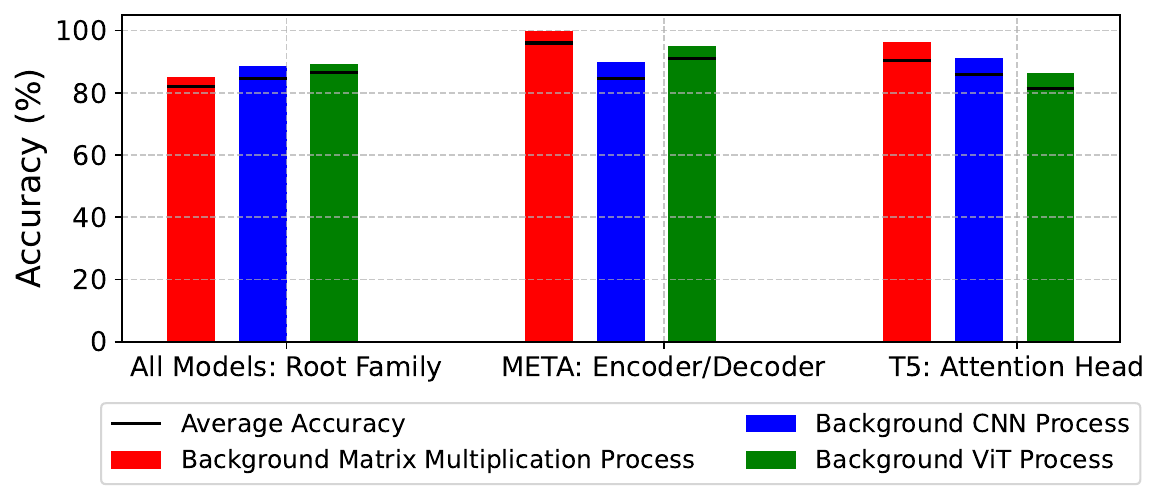}
   \caption{Prediction Model performance with different types of background process: Maximum Accuracy with Average as Dashed Line.\vspace{-4mm}}
\label{fig:Figure10}
\end{figure}

Additionally, we also want to verify our attack model against different noisy environments. In a realistic scenario, MIG instances often handle various categories of machine learning tasks simultaneously for specific users. 
To replicate this environment, we ran a transformer inference workload in parallel with large matrix multiplication, CNN-based image classification, and vision transformer inference, in three independent scenarios. The inclusion of these kind of resource hungry
tasks allowed us to assess how well our generalized prediction model performs in a complex, high-noise setting where GPU utilization is consistently high across multiple instances on a single physical device.
Under this setting, we observe (Figure~\ref{fig:Figure10}) a 4\% drop in root family identification accuracy when large matrix multiplication runs in parallel. However, this scenario outperformed the others in detecting \textit{META’s} encoder/decoder and \textit{T5’s} attention heads. Although performance varied across tasks in different scenarios, our prediction model consistently identified key architectural parameters of transformer models with high accuracy, even in noisy, MIG-enabled GPU environments.

\vspace{-4mm}

\section{Exploitation of Extracted Transformer Architecture}
\label{sec:attack}
In this section, we show how an adversary can exploit leaked architectural information—obtained through thermal and power side-channel analysis—to compromise a black-box transformer model. We highlight two possible consequences of such leakage. First, an adversary could build substitute models that closely mimic the behavior of the original model and use them for their own purposes. Second, the leaked information could enable more effective black-box transfer adversarial attacks, where adversarial examples crafted using a similar surrogate model are used to mislead the target model into making incorrect predictions.


\subsection{\textbf{Creation of Substitute Model}} In our first experiment, we built a substitute language transformer model replicating the MarianMT Helsinki architecture using side-channel-extracted details such as the number of encoders/decoders and attention heads. This model achieved a BLEU\footnote{BLEU score is an algorithm for evaluating the quality of text which has been machine-translated from one natural language to another.} score of 44.12 on the test set, indicating high-quality translations. As shown in Table~\ref{tab:surrogate_language_results}, it also attained a BERT\footnote{BERT score is an evaluation metric that compares candidate and reference sentences based on cosine similarity of contextual embeddings.} score of 93.53, closely matching the original model’s outputs when architectures aligned. 
These findings highlight that even partial architectural leakage can enable the creation of a shadow model that closely replicates the performance and outputs of the target model.

\begin{table}[ht]
\centering
\vspace{-1.5mm}
\caption{Substitute Language Transformer's Accuracy on Different Unknown Parameter Settings}
\label{tab:surrogate_language_results}
\resizebox{\columnwidth}{!}{%
\begin{tabular}{|c|c|c|c|c|c|}
\hline
\multirow{3}{*}{\textbf{Target Model}} & \multirow{3}{*}{\textbf{Substitute Model}} & \multirow{3}{*}{\textbf{Extracted Parameters}} & \multicolumn{2}{c|}{\textbf{Unknown Parameters}} & \multirow{2.5}{*}{\textbf{BERT}} \\
\cline{4-5}
 & & & \textbf{Embedding} & \textbf{Feed-forward} & \multirow{1.9}{*}{\textbf{Score}} \\
& & & \textbf{Dimension} & \textbf{Network} & \\
\hline
\multirow{2.5}{*}{Helsinki-NLP/} & \multirow{2.5}{*}{Helsinki-NLP/} & \multirow{2.5}{*}{Encoder/Decoder: 6} & 512 & 512 & 93.5 \\
\cline{4-6}
\multirow{2}{*}{opus-mt-de-en} & \multirow{2}{*}{opus-mt-de-en} & \multirow{2}{*}{Attention Head: 8} & 512 & 1024 & 93.08 \\
\cline{4-6}
 & & & 512 & 2048 & \textbf{93.53} \\
\hline
\end{tabular}%
}
\vspace{-4mm}
\end{table}

\subsection{\textbf{Black Box Transfer Adversarial Attack}} We further evaluated the implications of architectural leakage by conducting black-box transfer adversarial attacks on two target Vision Transformer (ViT) models from \textit{Google} family. For these models, we constructed substitute models using architectural information extracted via side channels and generated adversarial examples using the FGSM~\cite{goodfellow2015explainingharnessingadversarialexamples} and PGD~\cite{madry2019deeplearningmodelsresistant} methods, evaluating them against the corresponding target models. As shown in Table~\ref{tab:pre-trained_VIT_Attack_results}, both FGSM and PGD achieved an average attack success rate of 93\% for both substitute model architectures, indicating strong similarity with the target models. This highlights how leaked architectural information can enable highly effective adversarial attacks in black-box settings.

\begin{table}[h]
\centering
\vspace{-1mm}
\caption{Pre-Trained Vision Transformer's Classification Accuracy in Adversarial Attack Scenarios}
\label{tab:pre-trained_VIT_Attack_results}
\footnotesize
\begin{tabular}{|c|c|c|}
\hline
\textbf{Model} & \textbf{Attack} & \textbf{Success Rate (\%)} \\\hline
\multirow{2}{*}{Google/vit-base-patch16-224} & FGSM & 83.38 \\
                          \cline{2-3}
                          & PGD  & 94.87 \\\hline
\multirow{2}{*}{Google/vit-large-patch16-224} & FGSM & 98.63 \\
                          \cline{2-3}
                          & PGD  & 98.63 \\\hline
\end{tabular}
\vspace{-2mm}
\end{table}


\vspace{-2mm}
\section{Discussion}
\label{sec:discussion}

Our work goes beyond previous methods that mainly focused on CNNs using CPU side-channels or CUPTI-based GPU profiling. Transformers, unlike CNNs, have a more modular design with components like encoder-decoder blocks and attention heads, which create more visible patterns in power and thermal traces during inference. 
Moreover, while Transformers are typically larger than CNNs and may introduce more computational noise, their modular structure—such as repeated encoder/decoder blocks—creates stronger and more consistent patterns. This makes power and thermal side-channels more effective for detecting their behavior on GPUs even in noisy environments. Moreover, as CUPTI counters are often disabled or require root access on newer-generation GPUs, prior GPU-based techniques are becoming less practical. Power and thermal metrics, however, remain accessible through simple queries without elevated privileges. This makes our approach the first to successfully extract architectural information from transformer models on modern NVIDIA GPUs using side-channel signals alone.

We also observe that the most commonly downloaded  transformers on platforms like Hugging Face share a core set of architectural parameters—such as the number of layers, attention heads, and intermediate sizes—which we summarize in Table~\ref{tab:transformer_config} and \ref{tab:vision_transformer_config} of our paper. While differences may exist in datasets, embedding dimensions, or optimization methods, the core structure remains consistent. Therefore, our classification models are trained specifically to detect these known architectural patterns.



\vspace{-1mm}
\section{Mitigation and Conclusion}
\label{sec:mitigation_conclusion}

Power and thermal metrics are essential for GPU stability but can also leak sensitive model details. To mitigate this, we propose strategies on two fronts. On the \textit{application side}, combining model \textit{model pruning}~\cite{DBLP:conf/nips/KwonKMHKG22} and \textit{knowledge distillation}~\cite{DBLP:conf/cvpr/ChenCZZGT22} can generate obfuscated transformer variants with similar functionality, allowing random selection during inference to conceal architectural details. On the \textit{hardware or system side}, GPU vendors can restrict sensor access to admin users, exposing only key metrics like utilization and memory usage. In cases of high power or temperature, instance users can receive warnings, with minor throttling to maintain stability while limiting side-channel exposure.


In conclusion, this work investigates power and thermal side-channels as a novel approach to extract architectural details of transformer models on NVIDIA GPUs. These side-channels remains accessible without special privileges and cannot be virtualized, posing serious security risks. Using custom prediction models, we achieved close to 100\% accuracy in identifying key parameters of both custom and pre-trained transformers. For language models like \textit{T5} and \textit{Meta}, we accurately classified encoder/decoder layers and attention heads, while for vision transformers from \textit{Google}, \textit{Facebook}, and \textit{Microsoft}, we achieved 100\% accuracy in predicting attention head counts. Building on this extracted information, we successfully created substitute language transformer model with a BERT Score over 93, and achieves 93\% success in black-box transfer attacks on Vision Transformers, demonstrating the practical implications of such side-channel vulnerabilities. This highlights the urgent need to address side-channel threats to safeguard model architectures on GPUs.

\bibliographystyle{ieeetr}
{\small\bibliography{ICCAD_conference}}
\end{document}